\documentclass[final,5p,times,twocolumn,nopreprintline]{elsarticle}
\usepackage{amsmath,amssymb,amsfonts,dsfont}
\usepackage{graphicx}
\usepackage{slashed}
\usepackage{booktabs,tabulary}

\newcommand{\Fpi}{F_\pi}
\newcommand{\mpi}{M_{\pi}}
\newcommand{\mpii}{M_{\pi^0}}

\newcommand{\Order}{\mathcal{O}}
\newcommand{\muu}{m_u}
\newcommand{\md}{m_d}

\newcommand{\mN}{m_N}

\newcommand{\MeV}{\,\text{MeV}}
\newcommand{\GeV}{\,\text{GeV}}
\newcommand{\beq}{\begin{equation}}
\newcommand{\eeq}{\end{equation}}

\newcommand{\toright}[1]{\hspace*{\fill}{\footnotesize{#1}}}

\allowdisplaybreaks[1]

\begin{document}

\renewcommand{\theequation}{\arabic{equation}}

\begin{frontmatter}

\title{\toright{\textnormal{INT-PUB-16-006}}\\Remarks on the pion--nucleon $\sigma$-term}

\author[INT]{Martin Hoferichter}
\author[HISKP]{Jacobo Ruiz de Elvira}
\author[HISKP]{Bastian Kubis}
\author[HISKP,Julich]{Ulf-G.\ Mei{\ss}ner}

\address[INT]{Institute for Nuclear Theory, University of Washington, Seattle, WA 98195-1550, USA}
\address[HISKP]{Helmholtz--Institut f\"ur Strahlen- und Kernphysik (Theorie) and
   Bethe Center for Theoretical Physics, Universit\"at Bonn, D--53115 Bonn, Germany}
   \address[Julich]{Institut f\"ur Kernphysik, Institute for Advanced Simulation, 
   J\"ulich Center for Hadron Physics, JARA-HPC, and JARA-FAME,  Forschungszentrum J\"ulich, D--52425  J\"ulich, Germany}

\begin{abstract}
The pion--nucleon $\sigma$-term can be stringently constrained by the combination of analyticity, unitarity, and crossing symmetry with phenomenological information on the pion--nucleon scattering lengths. Recently, lattice calculations at the physical point have been reported that find lower values by about $3\sigma$ with respect to the phenomenological determination. 
We point out that a lattice measurement of the pion--nucleon scattering lengths could help resolve the situation by testing the values extracted from spectroscopy measurements in pionic atoms.  
\end{abstract}


\end{frontmatter}

\section{Introduction}

The pion--nucleon ($\pi N$) $\sigma$-term $\sigma_{\pi N}$ is a fundamental parameter of low-energy QCD. It measures the amount of the nucleon mass that originates from the up- and down-quarks, in contrast to the predominant contribution from the gluon fields generated by means of the trace-anomaly of the energy-momentum tensor. 
A precise knowledge of the $\sigma$-term has become increasingly important over the last years since it determines the
scalar matrix elements $\langle N|m_q\bar q q|N\rangle$ for $q=u,d$~\cite{Crivellin:2013ipa}, which, in turn, are crucial for the interpretation of dark-matter direct-detection experiments~\cite{Bottino:1999ei,Bottino:2001dj,Ellis:2008hf} and searches for lepton flavor violation in $\mu\to e$ conversion in nuclei~\cite{Cirigliano:2009bz,Crivellin:2014cta} in the scalar-current interaction channel.

Despite its importance, the value of $\sigma_{\pi N}$ has been under debate for decades. Phenomenologically, the $\sigma$-term can be extracted from $\pi N$ scattering by means of a low-energy theorem that relates the scalar form factor of the nucleon evaluated at momentum transfer $t=2\mpi^2$ to an isoscalar $\pi N$ amplitude at the Cheng--Dashen point~\cite{Cheng:1970mx,Brown:1971pn}, which unfortunately lies outside the region directly accessible to experiment. The necessary analytic continuation, performed in~\cite{Gasser:1988jt,Gasser:1990ce,Gasser:1990ap} based on the partial-wave analysis from~\cite{Koch:1980ay,Hoehler}, led to the classical value of $\sigma_{\pi N}\sim 45\MeV$~\cite{Gasser:1990ce}. Within the same formalism, this result was later contested by a new partial-wave analysis~\cite{Pavan:2001wz} that implied a significantly larger value $\sigma_{\pi N}=(64\pm 8)\MeV$.

Recently, a new formalism for the extraction of $\sigma_{\pi N}$ has been suggested relying on the machinery of Roy--Steiner equations~\cite{Ditsche:2012fv,Hoferichter:2012wf,Hoferichter:2015dsa,Hoferichter:2015tha,Hoferichter:2015hva}, a framework designed in such a way as to maintain analyticity, unitarity, and crossing symmetry of the scattering amplitude within a partial-wave expansion.
One of the key results of this approach is a robust correlation between the $\sigma$-term and the $S$-wave scattering lengths
\begin{align}
\label{sigma_piN_lin}
\sigma_{\pi N}&=(59.1\pm 3.1)\MeV + \sum_{I_s}c_{I_s}\big(a^{I_s}-\bar a^{I_s}\big),\\
c_{1/2}&=0.242\MeV\times 10^3\mpi,\qquad c_{3/2}=0.874\MeV\times 10^3\mpi,\notag
\end{align}
where the sum extends over the two $s$-channel isospin channels and $a^{I_s}-\bar a^{I_s}$ measures the deviation of the scattering lengths from their reference values extracted from pionic atoms
\begin{align}
\label{scatt_length_pionic}
\bar a^{1/2}&=(169.8\pm 2.0)\times 10^{-3}\mpi^{-1},\notag\\ 
\bar a^{3/2}&=(-86.3\pm 1.8)\times 10^{-3}\mpi^{-1}.
\end{align}
In this way, the main drawback of the formalism from~\cite{Gasser:1988jt,Gasser:1990ce}, the need for very precise input for a particular $P$-wave scattering volume, could be eliminated. In combination with the experimental constraints on the scattering lengths from pionic atoms, the low-energy theorem thus leads to a very stringent constraint~\cite{Hoferichter:2015dsa} 
\beq
\label{sigma_RS}
\sigma_{\pi N}=(59.1\pm 3.5)\MeV.
\eeq
Given that already $4.2\MeV$ of the increase originate from updated corrections to the low-energy theorem (thereof $3.0\MeV$ from the consideration of isospin-breaking effects), the net increase in the $\pi N$ amplitude compared to~\cite{Gasser:1990ce} adds up to about $10\MeV$, roughly half-way between~\cite{Gasser:1990ce} and~\cite{Pavan:2001wz}.  

While for a long time lattice calculations were hampered by large systematic uncertainties due to the extrapolation to physical quark masses, recently three calculations near or at the physical point appeared~\cite{Durr:2015dna,Yang:2015uis,Abdel-Rehim:2016won}, with results collected in Table~\ref{table:lattice}. All values lie substantially below the phenomenological value~\eqref{sigma_RS} 
(Table~\ref{table:lattice} also shows the significance in each case if all errors are added in quadrature). 
As we will argue in this Letter, this discrepancy should be taken very seriously as it suggests that the lattice $\sigma$-terms are at odds with experimental data on pionic atoms.
   
\begin{table}
\centering
\renewcommand{\arraystretch}{1.3}
\begin{tabular}{cccc}
\toprule
Collaboration & $\sigma_{\pi N}\, [\MeV]$ & Reference &\\\midrule
BMW & $38(3)(3)$ & \cite{Durr:2015dna} & $3.8\sigma$\\
$\chi$QCD & $44.4(3.2)(4.5)$ & \cite{Yang:2015uis} & $2.2\sigma$\\
ETMC & $37.22(2.57)\big(^{+0.99}_{-0.63}\big)$ & \cite{Abdel-Rehim:2016won} & $4.9\sigma$\\
\bottomrule
\end{tabular}
\caption{Lattice results for $\sigma_{\pi N}$. The last column gives the tension with~\cite{Hoferichter:2015dsa}, adding all errors in quadrature.
We do not attempt an average of the lattice results here.}
\label{table:lattice}
\renewcommand{\arraystretch}{1.0}
\end{table}

An analysis of flavor $SU(3)$ breaking suggests a $\sigma$-term closer to the small values obtained on the lattice 
(cf.\ the discussion in~\cite{Leutwyler:2015jga} and references therein): assuming 
violation of the OZI rule to be small, it should be not too far from the matrix element 
$\sigma_0 = (\muu+\md)/2\times\langle N|\bar uu+\bar dd-2\bar ss|N\rangle$, which can be 
related to the mass splitting in the baryon ground state octet and is usually found to be of 
the order of $35\MeV$~\cite{Gasser:1980sb,Borasoy:1996bx}.
However, significantly larger values have been obtained in the
literature when including effects of the baryon decuplet explicitly in
the loops, both in covariant and heavy-baryon
approaches~\cite{Alarcon:2012nr,Yao:2016vbz}, making it unclear how well the
perturbation series in the breaking of flavor $SU(3)$ behaves, and the
uncertainties difficult to quantify.

A similar puzzle emerged recently in a lattice calculation of $K\to\pi\pi$~\cite{Bai:2015nea}, which quotes a value of the isospin-$0$ $S$-wave $\pi\pi$ phase shift at the kaon mass $\delta^0_0(M_K)=23.8(4.9)(1.2)^\circ$, about $3\sigma$ smaller than the phenomenological result from $\pi\pi$ Roy equations~\cite{Colangelo:2001df,GarciaMartin:2011cn}.
A potential origin of this discrepancy could be that the strong $\pi\pi$ rescattering, known to be particularly pronounced in the isospin-$0$ $S$-wave, is not fully captured by the lattice calculation, given that the result for the isospin-$2$ phase shift $\delta^2_0(M_K)=-11.6(2.5)(1.2)^\circ$ is much closer to phenomenology. This explanation could be tested by a fully dynamical calculation of the corresponding scattering length $a^0_0$, which is predicted very accurately from the combination of Roy equations and Chiral Perturbation Theory~\cite{Colangelo:2000jc}, a prediction in excellent agreement with the available experimental information (see~\cite{Leutwyler:2015jga} for a review of the present situation). 
In the same way as $a^0_0$ provides a common ground where lattice, experiment, and dispersion theory can meet to resolve the discrepancy in the $\pi\pi$ case, a lattice measurement of the $\pi N$ scattering lengths could help clarify the $\sigma$-term puzzle. In this Letter we present our arguments why we believe this to be the case.   

\section{$\pi N$ scattering lengths from pionic atoms}

The linear relation~\eqref{sigma_piN_lin} between $\sigma_{\pi N}$ and the $\pi N$ scattering lengths proves to be a very stable prediction of $\pi N$ Roy--Steiner equations: while derived as a linear expansion around the central values~\eqref{scatt_length_pionic},
we checked the potential influence of higher terms by additional calculations on a grid around $\bar a^{I_s}$ with maximal extension of twice the standard deviation quoted in~\eqref{scatt_length_pionic} in each direction, with the result that also in this extended region quadratic terms are entirely negligible. The numbers for $c_{I_s}$ given in~\eqref{sigma_piN_lin} refer to this extended fit and therefore differ slightly from the ones given in~\cite{Hoferichter:2015dsa}. 
An additional check of the formalism is provided by the fact that if the scattering lengths from~\cite{Hoehler} are inserted, the lower $\sigma$-term from~\cite{Gasser:1990ce} is recovered. Irrespective of the details of uncertainty estimates, this behavior clearly demonstrates that the origin for the upward shift in the central value is to be attributed to the updated input for the scattering lengths.
The latter exercise also shows that the solution linearized around the pionic-atom reference point remains approximately valid in a much larger range of parameter space: for the scattering lengths from~\cite{Hoehler} it differs from the full solution by less than $2\MeV$.

In pionic atoms, electromagnetic bound states of a $\pi^-$ and a proton/deuteron core, strong interactions leave imprints in the level spectrum that are accessible to spectroscopy measurements~\cite{Gasser:2007zt}. In pionic hydrogen ($\pi H$) the ground state is shifted compared to its position in pure QED and acquires a finite width due to the decay to $\pi^0 n$ (and $n\gamma$) final states. The corresponding observables are therefore sensitive to the $\pi^-p\to\pi^-p$ and $\pi^-p\to\pi^0 n$ scattering channels. Although the width in pionic deuterium ($\pi D$) is dominated by $\pi^- d\to nn$, the level shift measures the isoscalar combination of $\pi^-p\to\pi^-p$ and $\pi^-n\to\pi^-n$, once few-body corrections are applied, and thus provides a third constraint on threshold $\pi N$ physics.
Experimentally, the level shifts have been measured with high accuracy at PSI~\cite{Strauch:2010vu,Hennebach:2014lsa}, and a preliminary value for the $\pi H$ width is reported in~\cite{Gotta:2008zza}.

At this level of accuracy a consistent treatment of isospin-breaking~\cite{Gasser:2002am,Meissner:2005ne,Hoferichter:2009ez,Hoferichter:2009gn} and few-body~\cite{Weinberg:1992yk,Beane:2002wk,Baru:2004kw,Lensky:2006wd,Baru:2007wf,Liebig:2010ki,Baru:2012iv} corrections becomes paramount if all three constraints are to be combined in a global analysis of $\pi H$ and $\pi D$~\cite{Baru:2010xn,Baru:2011bw}. 
In the isospin limit, the  $\pi N$ amplitude can be decomposed into two independent structures
\beq
T^{ba}=\delta^{ba}T^++\frac{1}{2}[\tau^b,\tau^a]T^-,
\eeq
where $a$ and $b$ refer to the isospin label of the incoming and outgoing pion, $\tau^a$ to isospin Pauli matrices, and $T^\pm$ to isoscalar/isovector amplitudes. Their threshold values are related to the $S$-wave scattering lengths by 
\beq
T^\pm\big|_\text{threshold}=4\pi\bigg(1+\frac{\mpi}{\mN}\bigg)a^\pm,
\eeq
where $\mpi$ and $\mN$ are the masses of pion and nucleon, and spinors have been normalized to $1$. Conventionally, the combined analysis of pionic atom data is not performed in terms of $a^+$, but using~\cite{Baru:2007ca} 
\beq
\label{atilde_def}
\tilde a^+ = a^+ + \frac{1}{4\pi\Big(1+\frac{\mpi}{\mN}\Big)}
\bigg\{\frac{4\Delta_\pi}{\Fpi^2}c_1-2e^2 f_1\bigg\}
\eeq
instead, where $\Delta_\pi=\mpi^2-\mpii^2$ denotes the pion mass difference, $\Fpi$ the pion decay constant, $e=\sqrt{4\pi\alpha}$, and $c_1$ and $f_1$ are low-energy constants that yield a universal shift in $a^+$ away from its isospin limit that cannot be resolved from pionic atoms alone. Moreover, we have defined particle masses in the isospin limit to coincide with the charged particle masses. 
The central values for the $s$-channel isospin scattering lengths~\eqref{scatt_length_pionic} have been obtained from such a combined analysis as follows~\cite{Hoferichter:2015hva}: first, we subtracted the contributions from virtual photons to avoid the presence of photon cuts, and second, we identified the $I_s=1/2,3/2$ channels from the physical $\pi^\pm p$ amplitudes
\begin{align}
a^{1/2}&=\frac{1}{2}\big(3a_{\pi^-p\to\pi^-p}-a_{\pi^+p\to\pi^+p}\big),\notag\\
a^{3/2}&=a_{\pi^+p\to\pi^+p}.
\end{align}
The main motivation for this convention is that $a_{\pi^-p\to\pi^-p}$ can be extracted directly from the $\pi H$ level shift without any further corrections, while $a_{\pi^+p\to\pi^+p}$ can be reconstructed from $a_{\pi^-p\to\pi^-p}$ and $\tilde a^+$ with minimal sensitivity to $a^-$ and thus the preliminary value for the $\pi H$ width. Of course, this convention has to be reflected in the precise form of the low-energy theorem for $\sigma_{\pi N}$~\cite{Hoferichter:2015dsa,Hoferichter:2015hva}, with uncertainties included in the error given in~\eqref{sigma_piN_lin}. 

To illustrate the tension between phenomenological and lattice determinations of $\sigma_{\pi N}$ it is most convenient to revert this change of basis by means of
\begin{align}
\label{SL_corr}
 a^{1/2}&=\tilde a^++2a^-+\Delta a^{1/2},\notag\\
 a^{3/2}&=\tilde a^+-a^-+\Delta a^{3/2},
\end{align}
where
\begin{align}
 \Delta a^{1/2}&=(-2.8\pm 1.3)\times 10^{-3}\mpi^{-1},\notag\\ 
\Delta a^{3/2}&=(-2.6\pm 0.7)\times 10^{-3}\mpi^{-1}.
\end{align}
The linear relation~\eqref{sigma_piN_lin} can then be recast as
\beq
\big(c_{1/2}+c_{3/2}\big)\tilde a^++\big(2c_{1/2}-c_{3/2}\big)a^-=C(\sigma_{\pi N}),
\eeq
where the right-hand side is given by
\begin{align}
C(\sigma_{\pi N})&=\sigma_{\pi N}-(59.1\pm 3.1)\MeV-\sum_{I_s}c_{I_s}\big(\Delta a^{I_s}-\bar a^{I_s}\big)\notag\\
&=\sigma_{\pi N}-(90.5\pm 3.1)\MeV.
\end{align}

\begin{figure}
\centering
\includegraphics*[width=\linewidth]{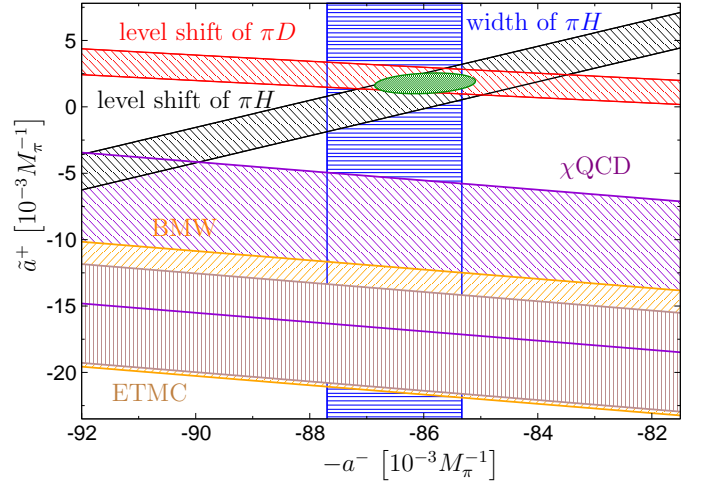}
\caption{Constraints on the $\pi N$ scattering lengths from pionic atoms (black: level shift in $\pi H$, blue: width of $\pi H$ ground state, red: level shift in $\pi D$) and from lattice $\sigma$-terms (orange: BMW~\cite{Durr:2015dna}, violet: $\chi$QCD~\cite{Yang:2015uis}, brown: ETMC~\cite{Abdel-Rehim:2016won}).}
\label{fig:bands_lattice}
\end{figure}

The corresponding bands in the $\tilde a^+$--$a^-$ plane are shown in Fig.~\ref{fig:bands_lattice}. As expected due to the isoscalar nature of the $\sigma$-term, the constraint from the lattice results is largely orthogonal to $a^-$, although non-linear effects in the Roy--Steiner solution generate some residual dependence on $a^-$ as well. The overall picture reflects the core of the discrepancy between lattice and phenomenology: while the three bands from the pionic-atom measurements nicely overlap, the lattice $\sigma$-terms favor a considerably smaller value of $\tilde a^+$.\footnote{In this context, it is also worth stressing that changing $a^{3/2}$ alone, where most of the difference between pionic atoms and~\cite{Hoehler} resides, is not an option: in doing so, one would infer, via the Goldberger--Miyazawa--Oehme sum rule~\cite{Goldberger:1955zza} that is sensitive to the isovector combination $a^-$, a value of the $\pi N$ coupling constant significantly too large compared to extractions from both nucleon--nucleon~\cite{deSwart:1997ep,Perez:2016aol} and pion--nucleon scattering~\cite{Arndt:2006bf}; see~\cite{Baru:2011bw}.} The exact significance again depends on if and how the three lattice measurements are combined, but in any case the fact remains that there is a disagreement with pionic-atom phenomenology around the $3\sigma$ level.   

\section{Lattice calculation of the $\pi N$ scattering lengths}

The discussion in the previous section makes it apparent that another independent determination of the $\pi N$ scattering lengths would imply additional information on $\sigma_{\pi N}$ that could help isolate the origin of the $\sigma$-term puzzle. Since a lattice calculation of $a^{I_s}$ would proceed directly in the isospin limit, we reformulate the relation~\eqref{sigma_piN_lin} accordingly. First,
we assume that the isospin limit would still be defined by the charged particle masses,\footnote{A similar analysis could be performed if the isospin limit were defined by the neutral pion mass. In this case, one would need to take the chiral isospin-limit expressions for the scattering lengths to adjust the pion mass from the charged to the neutral one, analogously to a chiral extrapolation.} but due to the absence of electromagnetic effects the corresponding scattering lengths as extracted from pionic atoms become
\begin{align}
\label{scatt_c}
 a_\text{c}^{1/2}&=a^{1/2}-\Delta a^{1/2}-\big(\tilde a^+-a^+\big)\notag\\
&=(178.8\pm 3.8)\times 10^{-3}\mpi^{-1},\notag\\
a_\text{c}^{3/2}&= a^{3/2}-\Delta a^{3/2}-\big(\tilde a^+-a^+\big)\notag\\
&=(-77.5\pm 3.5)\times 10^{-3}\mpi^{-1},
\end{align}
where we have used $c_1=-1.07(2)\GeV^{-1}$~\cite{Hoferichter:2015tha} and $|f_1|\leq 1.4\GeV^{-1}$~\cite{Fettes:2000vm,Gasser:2002am}.
The size of the shifts compared to~\eqref{scatt_length_pionic} is larger than one might naively expect from the chiral expansion, but the origin of the enhanced contributions is well understood: the bulk is generated from the term proportional to $4c_1\Delta_\pi/\Fpi^2$, see~\eqref{atilde_def}, which appears because the operator involving $c_1$ in the chiral Lagrangian generates a term proportional to the quark masses and thus, by the Gell-Mann--Oakes--Renner relation, to the neutral pion mass, which results in a large tree-level shift. The remainder is mainly due to a particular class of loop topologies, so-called triangle diagrams, which are enhanced by a factor of $\pi$ and an additional numerical factor.

In view of these effects one might wonder about the potential impact of $\Order(p^4)$ isospin-breaking corrections. However, both enhancement mechanisms will become irrelevant at higher orders simply due to the fact that the chiral $SU(2)$ expansion converges with an expansion parameter $\mpi/\mN\sim 0.15$ unless large chiral logs appear or additional degrees of freedom enhance the size of low-energy constants. This leaves as potentially large $\Order(p^4)$ corrections loop diagrams with low-energy constants $c_i$, which are numerically enhanced due to saturation from the $\Delta(1232)$, but at this order cannot appear in triangle-type topologies and therefore are not sufficiently enhanced to become relevant. Finally, similarly to $c_1$ at tree level, there is another artifact from the definition of the operator accompanying $c_2$, which is conventionally normalized to the nucleon mass in the chiral limit. At $\Order(p^4)$ this generates a quark-mass correction proportional to $c_1 c_2$ that renormalizes the aforementioned isospin-breaking correction involving $c_1$ by a factor $1+4c_2\mpi^2/\mN=1.27$, resulting in an additional shift in $a_\text{c}^{I_s}$ by $1.6$ units. Given that we do not have a full $\Order(p^4)$ calculation, we did not include this correction in the central values in~\eqref{scatt_c}, but, to stay conservative, in the quoted uncertainty as an estimate of the potential impact of higher-order terms.

\begin{figure}
\centering
\includegraphics*[width=\linewidth]{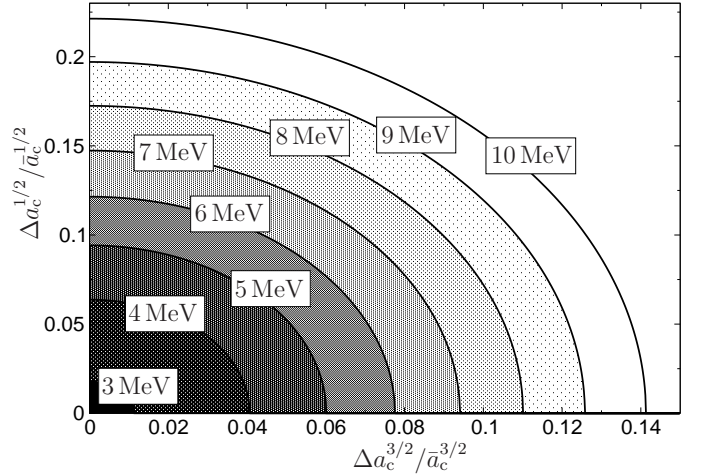}
\caption{Uncertainty in $\sigma_{\pi N}$ as a function of the relative accuracy in $a_\text{c}^{I_s}$.}
\label{fig:lattice_piN}
\end{figure}

If we finally rewrite~\eqref{sigma_piN_lin} in terms of $a_\text{c}^{I_s}$ in order to illustrate the impact of a lattice determination of the pion--nucleon scattering lengths on the $\sigma$-term, we obtain
\beq
\sigma_{\pi N}=(59.1\pm 2.9)\MeV + \sum_{I_s}c_{I_s}\big(a_\text{c}^{I_s}-\bar a^{I_s}_\text{c}\big),
\eeq
where the new reference values $\bar a^{I_s}_\text{c}$ refer to the central values given in~\eqref{scatt_c}. In this formulation the uncertainty even decreases slightly because the electromagnetic shift proportional to $f_1$ cancels to a large extent a similar correction in the low-energy theorem. 
The final uncertainty in $\sigma_{\pi N}$ for a given relative accuracy in the scattering lengths is shown in Fig.~\ref{fig:lattice_piN}.
For instance, if both isospin channels could be calculated at $[5\ldots10]\%$, one would obtain the $\sigma$-term with an uncertainty $[5.0\ldots 8.5]\MeV$.
We therefore see that to add conclusive information to the resolution of the $\sigma$-term puzzle by means of a lattice determination of the scattering lengths, a calculation at or below the $10\%$ level would be required. 
However, also more moderate lattice information may be helpful, e.g.\ in case one of the scattering lengths can be obtained more accurately than the other: as Fig.~\ref{fig:bands_lattice} suggests, also a single additional band could point towards significant tension with the very precise overlap region of the three pionic-atom experimental constraints.

\section{Conclusions}

In this Letter we highlighted the current tension between lattice and phenomenological determinations of the $\pi N$ $\sigma$-term. We argued that the puzzle becomes particularly apparent when formulated at the level of the $\pi N$ scattering lengths, which play a decisive role for the phenomenological value: a linear relation between the two scattering lengths of definite isospin and the $\sigma$-term allows one to reformulate any value for the latter as a constraint on the former, pointing towards a clear disagreement between lattice and pionic-atom data.
In a similar way as a direct lattice calculation of the isospin-$0$ $S$-wave $\pi\pi$ scattering length could help resolve a comparable discrepancy between lattice and Roy equations in $K\to\pi\pi$, we suggested that a lattice calculation of the $\pi N$ scattering lengths would amount to another independent determination of $\sigma_{\pi N}$ that could help identify the origin of the discrepancy.

\section*{Note added in proof}

While this paper was under review, another lattice calculation near the physical point appeared~\cite{Bali:2016lvx}. The quoted result $\sigma_{\pi N}=35(6)\MeV$ lies within the range of~\cite{Durr:2015dna,Yang:2015uis,Abdel-Rehim:2016won}.

\section*{Acknowledgements}

We thank Gilberto Colangelo, J\"urg Gasser, Heiri Leutwyler, and Martin J.~Savage for comments on the manuscript.
Financial support by
 the Helmholtz Virtual Institute NAVI (VH-VI-417),
the DFG (SFB/TR 16, ``Subnuclear Structure of Matter''), 
and the DOE (Grant No.\ DE-FG02-00ER41132) 
is gratefully acknowledged.
The work of UGM was supported in part by The Chinese Academy of Sciences 
(CAS) President's International Fellowship Initiative (PIFI) grant No.\ 2015VMA076.

\end{document}